\documentstyle[amssymb,preprint,aps]{revtex}

\begin{document}
\title{Missing Mass and the Acceleration of the Universe.\\
Is Quintessence the Only Explanation?}
\author{Sel\c{c}uk \c{S}. Bay\i n}
\address{Middle East Technical University\\
Department of Physics\\
Ankara TURKEY\\
bayin@metu.edu.tr}
\date{\today }
\maketitle

\begin{abstract}
\end{abstract}

Detailed observations of the temperature fluctuations in the microwave
background radiation indicate that we live in an open universe. From the
size of these fluctuations it is concluded that the geometry of the universe
is quite close to Euclidean. In terms Friedmann models, this implies a mass
density within 10\% of the critical density required for a flat universe.
Observed mass\ can only account for 30\% of this mass density. Recently, an
outstanding observation revealed that cosmos is accelerating. This motivated
some astronomers to explain the missing 70\% as some exotic dark energy
called quintessence. In this essay, we present an alternative explanation to
these cosmological issues in terms of the Friedmann Thermodynamics. This
model has the capability of making definite predictions about the geometry
of the universe, the missing mass problem, and the acceleration of the
universe in-line with the recent observations. For future observations, we
also predict where this model will start differing from the quintessence
models.

{\footnotesize \ This essay received an honorable mention in the Annual
Essay Competition of the Gravity Research Foundation for the year 2002--- Ed}%
..\newpage

\section{ Thermodynamics and Geometry}

Even though the left hand side of the Einstein's field equations is
constructed entirely from the metric tensor and its derivatives, the fact
that the right hand side (which describes the energy-momentum distribution)
also contains the metric tensor, indicates that matter and geometry are
interrelated in a deeper way. For a given geometry, Einstein's equations
could be used to obtain the total energy-momentum distribution of the
source. However, many different sources could be associated with a given
energy-momentum distribution. This indeterminacy about the details of the
source, which should be related to the information content of a given
geometry, immediately reminds us the entropy concept. In standard
statistical mechanics entropy is defined as proportional to the log of the
number of microstates that leads to the same macrostate. However, due to the
non-extensive nature of the self gravitating systems, even if we could find
a way to count the internal states of a given geometry, we could not expect
the corresponding 'curvature entropy' to be proportional to the log of this
number. A potential candidate may be the Tsallis' definition of entropy [1].

Another approach to the connection between geometry and thermodynamics could
be through the use of the second law, which states that the total entropy of
the universe can not decrease. However, due to the fact that it is the total
entropy that the second law is talking about, searching for a geometric
quantity that is an ever increasing function of time and identifying it as
the curvature entropy is not a reliable method. Besides, for non-extensive
systems the total entropy will not be a simple sum of its components [1].
Thus, making the contribution coming from the geometry even harder to
identify. Considering these difficulties, we have recently concentrated on
the thermodynamic side of this problem and argued that a system with finite
'curvature entropy' should also have a finite 'curvature temperature'. Being
homogeneous and isotropic, Friedmann geometries are ideal for searching this
connection, where the methods of equilibrium thermodynamics could still be
used [2-5].

Starting point of the Friedmann thermodynamics was the definition of the
'curvature temperature' as 
\begin{equation}
T=\alpha _{0}\left| \frac{k_{0}}{R(t)}\right| .
\end{equation}
$\alpha _{0}$ is a dimensional constant to be determined later and $R(t)$ is
the scale factor. $k_{0}^{2}/R(t)^{2}$ is proportional to the curvature
scalar of the constant time slices of the Friedmann geometry ( $k_{0}^{2}=0$
for critically open i.e. flat, $k_{0}^{2}=1$ for closed, and $k_{0}^{2}=-1$
for the open universes.). As expected from a temperature like property, (1)
is uniform throughout the system and also a three-scalar. With this new
information added to the Einstein's field equations, and for a 'local' (flat
spacetime) equation of state taken as 
\begin{equation}
P=\alpha \rho ,
\end{equation}
we were able to extract a 'global' equation of state that now incorporates
the effects of curvature as 
\begin{equation}
\ \rho _{open}(T,P)=-\frac{c_{0}^{2}}{8\pi }(3+\frac{1}{\alpha })T^{2}+\frac{%
P}{\alpha },\text{ \ \ \ \ where \ \ \ }c_{0}^{2}=\frac{4\pi ^{2}k^{2}c^{2}}{%
G\text{%
h\hskip-.2em\llap{\protect\rule[1.1ex]{.325em}{.1ex}}\hskip.2em%
}^{2}}\text{ , \ and }
\end{equation}
\begin{equation}
\rho _{closed}(T,P)=\frac{c_{0}^{2}}{8\pi }(3+\frac{1}{\alpha })T^{2}+\frac{P%
}{\alpha },
\end{equation}
for the open and closed models, respectively. These expressions, once
identified as the Gibbs energy densities could be used to derive all the
required thermodynamic properties of the system. Notice that $\rho $ and $P$
in (3,4) are no longer the same with their local values given in (1) [2,3].
They reduce to their local values in the flat spacetime limit i.e. $%
T\rightarrow 0$.

One remarkable consequence of this model is that one could now determine the
geometry of the universe by thermodynamic arguments. When we compare the two
geometries, we see that $\rho _{open}$ is always less than $\rho _{closed}.$
Thus, making it the more stable phase [2,3]. It is interesting that recent
observations on the inhomogeneities of the universal background radiation
also indicate that the universe is open, no matter how close it may be to
the critically open i.e. flat case [6,7].

In search for a justification of our definition of the curvature
temperature, we have studied Casimir energy in closed Friedmann models. By
taking the effective temperature of the Casimir energy as the curvature
temperature, we have identified the dimensional constant $\alpha _{0}$ as $\ 
\frac{1}{2\pi }\frac{\text{%
h\hskip-.2em\llap{\protect\rule[1.1ex]{.325em}{.1ex}}\hskip.2em%
c}}{k}.$ Later, by using the concept of local thermodynamic equilibrium, we
have extended our definition of curvature temperature to the sufficiently
slowly varying but otherwise arbitrary spacetimes [4]. When this definition
was used for spherically symmetric stars, we have shown that in the black
hole limit, the curvature temperature at the surface of the star reduces to
the Hawking temperature, precisely.

\section{Changes in the Local Equation of State and the \ Friedmann
Thermodynamics}

A large class of phase changes in the local matter distribution, which
includes the transition from the radiation era to the matter era could be
described as 
\begin{equation}
P=\alpha _{1}\rho \text{ \ }\rightarrow \text{ }P=\alpha _{2}\rho \text{ .}
\end{equation}
Such transitions, aside from a change in the amount of deceleration, do not
have any interesting consequences within the context of standard Friedmann
models. However, considered in the light of Friedmann thermodynamics, they
offer new answers to some of the basic issues of cosmology.

We will now concentrate on the beginning of the galaxy formation era, where
the $\alpha $ value of the universe is expected to decrease. This follows
from the fact that at the onset of the galaxy formation, some of the gas in
the universe will be immobilized, thus giving less pressure for the same
mean density. For $P=\alpha \rho $, and an open universe the Gibbs energy
density was given in (3). For the transition $\alpha _{1}\rightarrow \alpha
_{2},$ the difference between them could be written as 
\begin{equation}
\rho _{open,\alpha _{2}}(T,P)-\rho _{open,\alpha _{1}}(T,P)=\frac{%
c_{0}^{2}(\alpha _{2}-\alpha _{1})}{8\pi \alpha _{1}\alpha _{2}}[T^{2}-\frac{%
8\pi }{c_{0}^{2}}P].
\end{equation}
The two surfaces intersect along the curve 
\begin{equation}
T_{c}^{2}=\frac{8\pi }{c_{0}^{2}}P_{c}\text{ \ .}
\end{equation}
We could use the curvature temperature at the onset of the galaxy formation
era as the critical temperature $T_{c},$ and obtain $P_{c}$ from the above
relation. In ordinary phase transitions critical temperature is usually
defined with respect to the constant atmospheric pressure. In our case, at
the critical point both phases are expected to coexist, thus it is natural
to expect $P_{c}$ to lie somewhere in between the pressures just before the
transition has started, and after it has completed. In this regard, due to a
reduction in the local pressure, we expect $T^{2}-\frac{8\pi }{c_{0}^{2}}P<0$
before the critical point is reached, and $T^{2}-\frac{8\pi }{c_{0}^{2}}P>0$
after the transition is completed . Considering that $\alpha _{2}-\alpha
_{1}<0$, we could conclude that $\rho _{open,\alpha _{1}}(T,P)$ , and $\rho
_{open,\alpha _{2}}(T,P)$ are the stable phases before and after the
critical temperature, respectively.

\section{Dark Energy or the Missing Mass}

Now let us now see what new insights that this model contribute to
cosmology. Enthalpy density corresponding to the global equation of state
(3) could be written as 
\begin{equation}
h(s,P)=\frac{8\pi }{4c_{0}^{2}}(3+\frac{1}{\alpha })^{-1}s^{2}+\frac{P}{%
\alpha },
\end{equation}
where $s$ is the entropy density. During the phase transition ($\alpha
_{1}\rightarrow \alpha _{2})$ change in the enthalpy density could be
written as 
\begin{eqnarray}
\Delta h(s,P) &=&\frac{8\pi }{2c_{0}^{2}}(3+\frac{1}{\alpha })^{-1}s\Delta s+%
\frac{1}{\alpha }\Delta P,\text{ and} \\
\Delta h(s,P) &=&T\Delta s+\frac{1}{\alpha }\Delta P.
\end{eqnarray}
At constant pressure $\Delta h(s,P)$ gives us the energy density needed for
this phase transition. In ordinary phase transitions this energy would be
absorbed from a heat bath at constant temperature. In our case, since the
universe is a closed system, it could only come from within the system.
Calling this energy density $q_{c},$ we obtain it as 
\begin{equation}
q_{c}=\frac{2c_{0}^{2}}{8\pi }T_{c}^{2}\frac{(\alpha _{1}-\alpha _{2})}{%
\alpha _{1}\alpha _{2}},\text{ \ \ \ where}
\end{equation}
\begin{equation}
T_{c}=\frac{1}{2\pi }\frac{\text{%
h\hskip-.2em\llap{\protect\rule[1.1ex]{.325em}{.1ex}}\hskip.2em%
c}}{k}\frac{1}{R_{c}}.
\end{equation}
$R_{c}$ is the scale factor of the universe at the time of the transition,
and $q_{c}$ is the energy spent (used) by the system (universe) to perform
the above phase transition, which is required by the entropy criteria. In
the energy budget of the universe, this energy would show up as missing with
respect to the critically open (flat) case. To find how this energy would be
observed today, we use the scaling property of $q_{c}$, to obtain 
\begin{equation}
q_{now}=\frac{2}{8\pi }\frac{c^{4}}{G}\frac{(\alpha _{1}-\alpha _{2})}{%
\alpha _{1}\alpha _{2}}\frac{1}{R_{now}^{2}}.
\end{equation}
Cosmic microwave background radiation data is sometimes used to claim that
the geometry of the universe is flat (critically open). However, all it
actually says is that no matter how close it may be to being flat, the
geometry is open. The huge difference between the two cases is usually
ignored. For the universe to be flat, its density must be tuned to the
critical density with infinite precession. From the Boomerang data, all one
could conclude is that the density of the universe is within 10\% of the
critical value [6,7]. Dynamical mass measurments indicate that the matter
content of the universe [radiation 0.005\%, ordinary visible matter
(baryonic) 0.5\%, ordinary non-luminous matter (baryonic) 3.5\%, and exotic
dark matter (WIMPS $\left[ 8,9\right] $; Basically non-baryonic but observed
through their gravitational effects.) 26\%] only adds up to 30\% of the
critical density. The rest is usually declared either as 'missing', or
exotic dark energy (quintessence) [10,11]. In the light of recent
measurments we could take the Hubble constant in the range [55-75]km.sec$%
^{-1}$.Mpc$^{-1}$[12]$.$ This gives the range of the critical density $(%
\frac{3H_{0}^{2}}{3\pi G})$ as 
\begin{equation}
\rho _{critical}\in \lbrack 6.05\text{ x}10^{-30},\text{ }1.44\text{ x}%
10^{-29}]gm/cc,
\end{equation}
which indicates that the range of the missing mass density is; 
\begin{equation}
\rho _{\text{missing}}\in \lbrack 4.23\text{ x}10^{-30},\text{ }1.01\text{ x}%
10^{-29}]gm/cc.
\end{equation}
Since the baryonic matter is the main source of pressure during the matter
era, we could take its average equation of state as the ideal gas law [13]
i.e. 
\begin{equation}
P=\frac{2}{3}u,\text{ \ \ \ \ where }u\text{ is energy density of baryons.}
\end{equation}
However, even though the baryonic matter is the main source of pressure, it
is not the main source of matter. In the light of recent observations it as
seen that baryonic matter only adds up to 4-5\% of the critical density
[6-11]. In terms of the total energy density we could now calculate the
ranges of the effective $\alpha _{1}$ and $\alpha _{2}$values as 
\begin{equation}
\alpha _{1}\in \frac{2}{3}[\frac{4}{100},\frac{5}{100}],\text{ \ \ \ }\alpha
_{2}\in \frac{2}{3}[\frac{3.5}{100},\text{ }\frac{4.5}{100}].
\end{equation}
We have assumed that roughly 10\% of this baryonic matter is converted into
luminous matter, thus decoupling from the expansion of the universe $\left[
10\right] $. The rest of the matter is considered as weakly interacting dark
matter $\left[ 8,9\right] $. We could now calculate the drop in the value of 
$\alpha $ to be in the range 
\begin{equation}
\frac{(\alpha _{1}-\alpha _{2})}{\alpha _{1}\alpha _{2}}\in \left[ 3.33,%
\text{ }5.36\right] .
\end{equation}
The value of $R_{now}$ is not directly observable but we could take it
roughly as the radius of the visible universe [14]: 
\begin{equation}
R_{now}\approx 2\text{ x}10^{28}cm.=2\text{ x}10^{10}ly.
\end{equation}
Using these, we could calculate $q_{now}$ (13) as 
\begin{equation}
q_{now}=\ \left[ 8.89\text{ x}10^{-30},\text{ }1.43\text{ x }10^{-29}\right]
gm/cc.
\end{equation}
This result for $q_{now}$ is now well in the range given in (15). Certainly
this energy does not disappear from the universe, but it is needed for the
phase transition, and it is used for it [13].

\section{Where did the Missing Mass Go -\ Acceleration?}

To see how $q_{now}$ is spent, we write the free energy density and its
change as 
\begin{eqnarray}
f(T,v) &=&-\frac{c_{0}^{2}}{8\pi }(3+v)T^{2}, \\
\Delta f(T,v) &=&-\frac{c_{0}^{2}}{8\pi }2T(3+v)\Delta T-\frac{T^{2}c_{0}^{2}%
}{8\pi }\Delta v, \\
\Delta f(T,v) &=&-s\Delta T-P\Delta v.
\end{eqnarray}
For constant temperature processes, $\Delta f(T,v)$ would usually give the
work done by the system on the environment through the action of a boundary.
Since we have a closed system, this work is done by those parts of the
system expanding under its own internal pressure: 
\begin{equation}
w_{c}=P_{c}\Delta v=\frac{c_{0}^{2}T_{c}^{2}}{8\pi }\frac{(\alpha
_{1}-\alpha _{2})}{\alpha _{1}\alpha _{2}},
\end{equation}
where $w_{c}$ denotes the work done by the system at the time of the
transition. Now at the critical point, $q_{c}$ amount of energy is used from
the system, while $w_{c}$ amount of it is used to do work to increase the
specific volume. In a closed system we expect these two terms to cancel each
other. However, as opposed to the usually studied systems in thermodynamics,
where changes take place infinitesimally slowly, our system is dynamic i.e.
Universe does not stop and go through this phase transition infinitesimally
slowly. As a result, we should also take into account the change in the
kinetic energy of the expansion. Hence, change in the internal energy should
be written as 
\begin{equation}
-q_{c}+w_{c}+\Delta (K.E.)_{c}=0.
\end{equation}
This implies 
\begin{equation}
\Delta (K.E.)_{c}=\text{ }\frac{c^{4}}{8\pi GR_{c}^{2}}\frac{(\alpha
_{1}-\alpha _{2})}{\alpha _{1}\alpha _{2}}>0\text{ \ \ (in ergs/cc).}
\end{equation}
This is the amount of energy that has gone into increasing the kinetic
energy of the expansion. In this model $q_{c}$ amount of energy (density)
has been used (or converted) within the system. Part of it has gone into
work to increase the specific volume, while the rest is used\c{c}n to
increase the kinetic energy of the expansion. Why should the universe go
through all this trouble? Basically, for the same reason that water starts
boiling when the critical temperature is reached i.e. to increase its
entropy.

Notice that all the arguments appearing in this paper are local and
independent of global topological concerns, which are to be addressed
separately. In principle, an infinite homogeneous universe with infinite
amount of matter, created at once and has finite age is as pathological as
any other infinity in our theories [13].

Let us finally estimate the change in the Hubble constant [13]. Taking the
average kinetic energy density of the expansion, in a volume large enough
for the effects of curvature to be important as $K.E.=\frac{1}{2}\rho 
\stackrel{.}{R}^{2},$ we could write $\Delta (K.E.)=\rho R^{2}H^{2}\frac{%
\Delta H}{H}.$ At the time of the transition (actually after it has been
completed) this is equal to $\left( 26\right) $ thus, we could write 
\begin{equation}
(\frac{\Delta H}{H})_{c}=4.83\text{ x10}^{47}\frac{1}{\rho
_{c}R_{c}^{4}H_{c}^{2}}\frac{(\alpha _{1}-\alpha _{2})}{\alpha _{1}\alpha
_{2}}.
\end{equation}
In this model effect of this phase transition will show up as a
discontinuity in the slope of the Hubble diagram $\left[ 15\right] ,$
roughly given by the amount given in (27). Location of this discontinuity
could be estimated roughly to be in the range $(0.54,0.91)$. In other words,
at the beginning cosmos was expanding slower while decelerating. During the
formation of galaxies it briefly accelerates for the reasons explained
above, and currently it is expanding faster but still decelerating. Recent
data with $0.5<$ $z<0.9$ just began to display the change in the Hubble
[15-17]. We expect that the deceleration will re-appear as more data with
redshifts $z\gtrsim 1.0$ is gathered. These galaxies will be among the very
first galaxies formed in the universe thus, still showing the dynamics of
the pre-galaxy formation era. For the nearby galaxies with redshifts $%
z\lesssim 0.5$ , we expect to see the dynamics after the transition is
completed. Thus galaxies with redshifts close to the upper end of this
interval should also show some deceleration. This is in contrast to the
predictions of the quintessence models, where the acceleration continues
forever at an ever increasing pace $\left[ 10,11\right] $.

\newpage

\section{References}

[1]R. Salazar, and R. Toral, Physica, {\bf A290},159 (2001).

[2]S. Bayin, Ap. J.{\bf , 301}, 517 (1986).

[3]S. Bayin, Gen.Rel.Grav., {\bf 19}, 899{\bf \ (}1987{\bf ).}

[4]S. Bayin, Gen.Rel.Grav., {\bf 22},179{\bf \ (}1990{\bf ).}

[5]S. Bayin, Gen.Rel.Grav., {\bf 26},{\bf \ }951 (1994).

[6]J. Silk, Physics World, {\bf 13}, no.6, pg.23 (2000).

[7]C. Baugh, and C. Frenk, Physics World, {\bf 12}, no.5, pg.25 (1999).

$\left[ 8\right] $A. Taylor, and J. Peacock, Physics World, {\bf 14}, no.3,
pg.37 (2001).

$\left[ 9\right] $N. Smith, and N. Spooner, Physics World, {\bf 13}, no.1,
pg.23 (2000).

[10]J. P. Ostriker, and P.J. Steinhardt, Sci. Am., Jan., pg.37 (2001).

[11]R. R. Caldwell, and P. J. Steinhardt, Physics World, {\bf 13}, no.11,
pg.31 (2000).

[12]R. Ellis, Physics World, {\bf 12}, no7, pg.19 (1999).

[13]S. Bayin, Submitted to ApJ{\bf \ (}2002{\bf ).}

[14]S. D. Landy, Sci. Am., June, pg.30 (1999).

[15]C. J. Hogan, R. P. Kirshner, and N. B. Suntzeff, Sci. Am., Jan., pg.28
(1999).

[$16]$M. Rees, Sci.\ Am., December, pg.44 (1999).

$\left[ 17\right] $D.Goldsmith, Science, {\bf 276}, 37 (1997).

\bigskip\ 

\end{document}